\documentclass[twocolumn]{article}

\usepackage{geometry}
\usepackage{doi}

\usepackage{etoolbox}

\patchcmd{\thebibliography}
  {\settowidth}
  {\setlength{\itemsep}{0pt}%
   \setlength{\parskip}{0pt}%
   \settowidth}
  {}{}

\usepackage{titlesec}
\titleformat*{\section}{\large\bfseries}
\titleformat*{\subsection}{\normalsize\bf}

\usepackage{sectsty}
\allsectionsfont{\raggedright}
\sectionfont{\raggedright\large\selectfont}        
\subsectionfont{\raggedright\normalsize\selectfont}  
\subsubsectionfont{\raggedright\itshape\normalsize\selectfont} 

\usepackage{graphicx}

\usepackage{hyperref}
\hypersetup{colorlinks=true,linkcolor=blue,citecolor=blue,urlcolor=blue}
\usepackage{cite}

\usepackage{bm}
\usepackage{amsmath}
\usepackage{amssymb}
\usepackage{times}
\usepackage[labelfont=bf]{caption}

\usepackage{abstract}

\begin{document}

\title{\textbf{Ptychographic estimation of qudit states encoded in the angular position and orbital angular momentum of single photons}}

\author{\fontsize{11pt}{25pt}\selectfont\textbf{A. M. da Costa\, and L. Neves}
\vspace{1mm} \\
\textit{\normalsize{Departamento de F\'isica, Universidade Federal de Minas Gerais, Belo Horizonte, MG, Brazil}} \\
}

\date{}

\twocolumn[
  \begin{@twocolumnfalse}
    \maketitle
\vspace{-7mm}
    \begin{abstract}

\noindent Ptychography is a computational imaging technique mainly used in optical and electron microscopy. Its quantum analogue was recently introduced as a simple method for estimating unknown pure quantum states through projections onto partially overlapping subspaces, each one followed by a projective measurement in the Fourier basis. In the end, an iterative algorithm estimates the state from the collected data. Here, we theoretically describe how to implement this method for $D$-dimensional qudit states encoded in the angular position and orbital angular momentum (OAM) of single photons. For this purpose, we define the qudit by discretizing the spatial profile of a photon in cylindrical coordinates, using an array of $D$ angular slits symmetrically distributed in the transverse plane. To apply ptychography to this encoding, we show that the intermediate projections will be carried out by simple binary spatial filters in the angular paths, while the measurement in the Fourier basis will be performed by postselecting $D$ OAM modes compatible with the quantum Fourier transform of the path basis. We illustrate the effectiveness of this scheme through simulations and discuss its experimental feasibility. \\

\vspace{-3mm}
\vspace{5mm}
    \end{abstract}
  \end{@twocolumnfalse}
]

\section{Introduction}

Estimating the unknown state describing an ensemble of quantum systems is a fundamental task in quantum information processing \cite{Banaszek2013,TeoBook}. Standard techniques rely on performing projections in a complete set of incompatible bases \cite{Wootters1989}, which becomes increasingly complex as the state-space dimension grows \cite{James2001,Thew2002}.  Considerable efforts have been devoted to reducing this complexity by exploiting prior information about the state to be determined. One approach in this regard involves the prior knowledge that the state is pure. Several methods have been suggested recently \cite{Goyeneche2015,Carmeli2016,Stefano2017,Stefano2019,Zambrano2020,Fernandes2019}, demonstrating that for estimating pure states of a \emph{qudit}, a $D$-dimensional quantum system, the required number of measurement outcomes scales linearly with $D$, in contrast to the $D^2$ scaling of standard schemes. For several physical platforms, current technology enables the preparation of quantum states with high purity. These methods are expected to estimate them with high fidelity, using significantly fewer resources.

Among the pure-state estimation methods, here we will address the one proposed in Ref.~\cite{Fernandes2019} as an analogue of the computational imaging technique known as \emph{ptychography} \cite{Faulkner04,Rodenburg04}. In this technique, primarily used in optical \cite{Thibault2008} and electron \cite{Humphry2012} microscopy, the complex-valued transmission function of an object is estimated computationally. This is achieved through an iterative algorithm that analyzes multiple diffraction patterns generated by shifting a coherent illumination across partially overlapping sections of the object. In quantum ptychography, an unknown pure state is subjected to multiple projections onto partially overlapping subspaces of the state space (analogous to the shifting illuminations), each one followed by a projective measurement in the Fourier basis (analogous to the diffraction patterns). The measurement outcomes are fed into an iterative phase-retrieval algorithm that estimates the complex amplitudes characterizing the state. While typical approaches for pure-state estimation require projective measurements in a variety of bases \cite{Goyeneche2015,Carmeli2016,Stefano2017,Stefano2019,Zambrano2020}, which can be experimentally challenging, the ptychographic method employs a single measurement basis. The diversity of the data is provided by the intermediate projections, which are straightforward to implement, as we will see later.

Recently, Fernandes \textit{et al.}\ \cite{Fernandes2020} experimentally demonstrated the ptychographic estimation of qudit states encoded in the linear transverse position and momentum of photons. Here, we theoretically study the angular version of this protocol, i.e., the ptychographic estimation of qudit states encoded in the angular position and orbital angular momentum (OAM) of single photons, which are an important resource for high-dimensional quantum information processing \cite{Erhard2018}. Photonic angular qudits are generated by discretizing the spatial profile of a photon, using an array of $D$ angular slits \cite{Jha2010,Puentes2020,Zhang2022}. To apply quantum ptychography to this encoding, we show that the partially overlapping projections will be carried out by simple binary spatial filters blocking $D-\lceil D/2\rceil$ angular modes. In addition, it is demonstrated that the measurement in the Fourier basis will be performed by postselecting $D$ OAM modes compatible with the quantum Fourier transform of the angular basis. We also provide the optimal way to select these modes from the infinite possibilities. To illustrate the effectiveness of our proposal, we simulate the estimation of five- and twelve-dimensional angular qudit states. The feasibility of the method is discussed in light of the currently available optical tools to manipulate and measure angular and OAM modes.

This article is organized as follows. Section~\ref{sec:Ptychography} outlines the ptychographic method. Sections~\ref{sec:AngularQudit} and~\ref{sec:PtyAngQudit} describe photonic angular qudit states and their ptychographic estimation, respectively. Conclusions are given in Section~\ref{sec:Conclusion}.


 \section{Ptychographic estimation of qudit states}
 \label{sec:Ptychography}

\begin{figure}[t]
\centering\includegraphics[width=1\columnwidth]{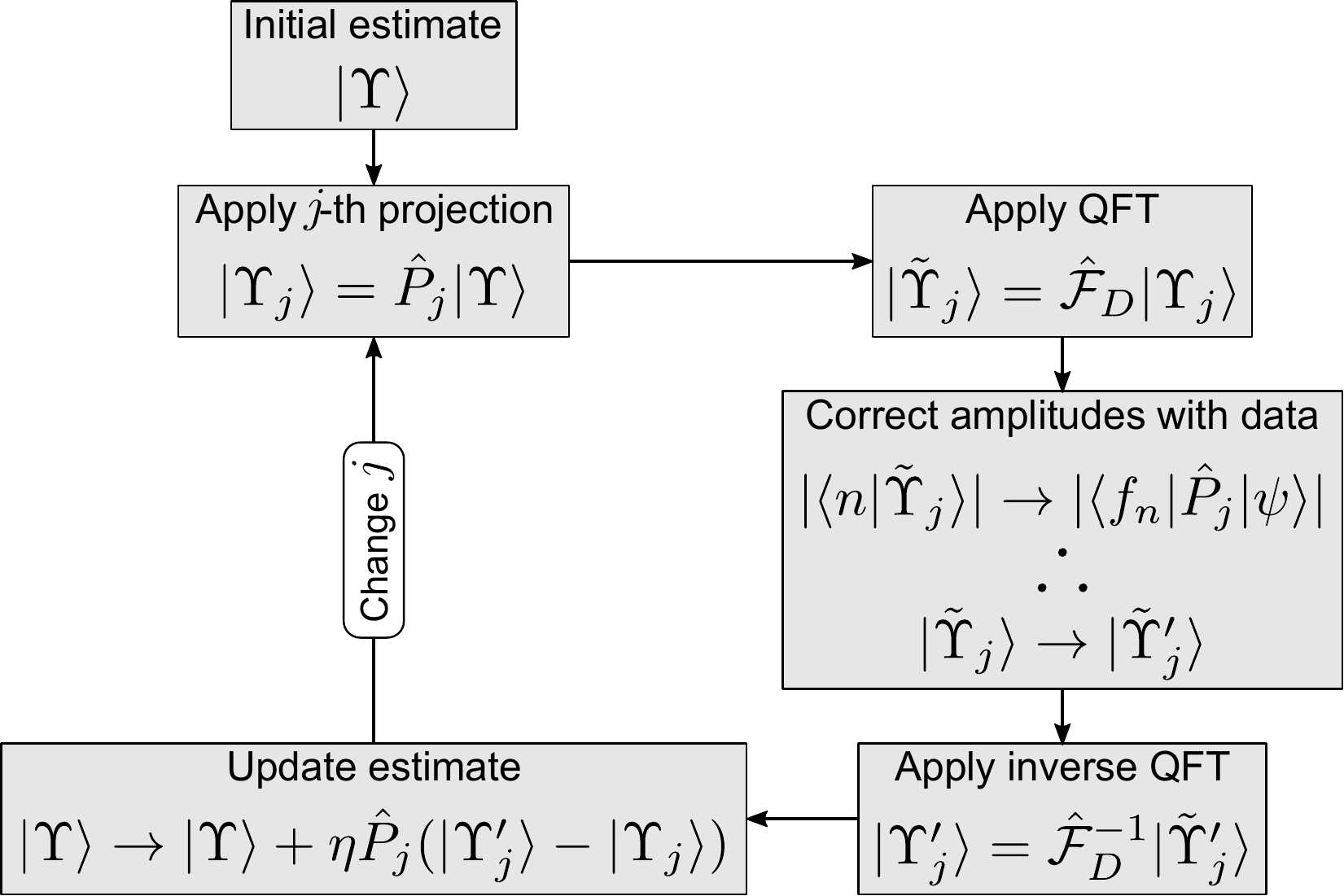}
    \caption{PIE algorithm: an initial random state, $|\Upsilon\rangle$, goes through the steps shown in the diagram where each ptychographic projector, $\hat{P}_j$, and its corresponding data, $\{|\langle f_n|\hat{P}_j|\psi\rangle|\}_{n=0}^{D-1}$, is used to update the estimate. See text for more details.} 
    \label{fig:PIE}
\end{figure}

Let $\{|n\rangle\}_{n=0}^{D-1}$ denote an orthonormal basis spanning a $D$-dimensional Hilbert space, $\mathcal{H}_D$, of a given quantum system. An arbitrary pure state of this system can be written in this basis as
\begin{equation}
|\psi\rangle=\sum_{n=0}^{D-1}C_n|n\rangle,
\end{equation}
where $\sum_n|C_n|^2=1$. The complex coefficients $\{C_n\}$ fully characterize the state and can be estimated by the ptychographic method as follows \cite{Fernandes2019}. First, one of $J$ rank-$\lceil D/2\rceil$ projectors given by
\begin{equation}\label{eq:PtyProjectors}
    \hat{P}_j = \sum_{k=0}^{\lceil D/2\rceil-1}|k \oplus s_j\rangle \langle k \oplus s_j|,
\end{equation}
is applied on the input state $|\psi\rangle$, where $j=0,\ldots,J-1$, $\oplus$ denotes addition modulo $D$, and $s_j$ is a nonnegative integer (see discussion below). The output state $|\psi_j\rangle=\hat{P}_j|\psi\rangle$ is then subjected to a projective measurement in the Fourier basis: 
\begin{equation}  \label{eq:FourierBasis}
    \hat{\mathcal{F}}_D|n\rangle \equiv|f_n\rangle = \dfrac{1}{\sqrt{D}}\sum ^{D-1}_{k=0}e^{2\pi ink/D} |k\rangle,
\end{equation}
where $\hat{\mathcal{F}}_D$ is the quantum Fourier transform (QFT) acting on $\mathcal{H}_D$. This procedure is repeated for each projector $\hat{P}_j$ and the generated data, proportional to $\{|\langle f_n|\hat{P}_j|\psi\rangle|^2\}_{n=0}^{D-1}$, are recorded. The state is then estimated by an iterative phase-retrieval algorithm, known as the ptychographic iterative engine (PIE) \cite{Faulkner04,Rodenburg04}. An initial random state, $|\Upsilon\rangle$, goes through the steps shown in the PIE diagram of Fig.~\ref{fig:PIE}. The $j$th ptychographic projector $\hat{P}_j$ and the corresponding data are used to update the initial estimate; $\eta\in(0,2]$ is a feedback parameter that controls the step-size of the update. A PIE iteration is completed after $J$ cycles through the closed loop, when each projector and the associated data are used once. At each iteration, the trace distance between the normalized current ($|\Upsilon_\textrm{curr}\rangle$) and updated ($|\Upsilon_\textrm{updt}\rangle$) estimates is computed \cite{NielsenBook}:
\begin{equation}   \label{eq:trace_distance}
\Delta = \sqrt{1-|\langle\Upsilon_\textrm{curr}|\Upsilon_\textrm{updt}\rangle|^2},
\end{equation}
where $\Delta\in[0,1]$. $\Delta$ is expected to decrease as the algorithm progresses, since the estimates become less distinguishable. The algorithm may be finished after a fixed number of PIE iterations or when $\Delta$ falls below a predefined threshold. Once complete, a pure state, pending normalization, will be provided.

As shown in Ref.~\cite{Fernandes2019}, the number of ptychographic projectors given by (\ref{eq:PtyProjectors}) is flexible, as long as each $\hat{P}_j$ has a partial overlap with at least one partner (e.g., $0<\textrm{Tr}\{\hat{P}_j\hat{P}_{j'}\}/\lceil D/2\rceil <1$) and all levels of $\mathcal{H}_D$ are addressed at least once. A given set is defined by the $J$-entry vector $\mathbf{s}_J=(s_0,\ldots,s_{J-1})$, where $J$ may vary between $D$ and a fixed value ($<D$) regardless of the qudit dimension. For instance, using $J=D$ projectors with $\mathbf{s}_D=(0,\ldots,D-1)$, one generates $D^2$ measurement outcomes, which is an overcomplete dataset to estimate pure states. On the other hand, using $J=5$ projectors with $\mathbf{s}_5=\{j\lfloor D/5\rfloor\}_{j=0}^4$, the state is estimated with $5D$ measurement outcomes, which is comparable to other pure-state estimation methods \cite{Goyeneche2015,Carmeli2016,Stefano2017,Stefano2019,Zambrano2020}.

\section{Photonic angular qudits}
\label{sec:AngularQudit}
\subsection{Discretizing the angular position of a photon}
\label{subsec:AngularQudit2}

Let us consider a source that emits single monochromatic photons with spatially homogeneous polarization, which propagate paraxially along a  direction defined as the $z$ axis. The state of these photons can be described by the spatial probability amplitude in the transverse plane, $\Psi(\mathbf{r}_{\perp})$, where $\mathbf{r}_{\perp}$ represents the position coordinate in this plane. In the case of a pure state, we will then have
\begin{equation}\label{eq:estadoinicial}
    |\Psi\rangle = \int d \mathbf{r}_\perp  \Psi(\mathbf{r}_\perp) |\mathbf{r}_\perp\rangle,
\end{equation}
where $\int d\mathbf{r}_\perp |\Psi({\mathbf{r}_\perp})|^{2}  = 1$. Let us further consider that this probability amplitude, in cylindrical coordinates, is separable into radial ($\rho$) and azimuthal ($\phi$) coordinates, i.e.,
\begin{equation}\label{eq:separability}
\Psi(\mathbf{r}_\perp) =  \Psi(\rho,\phi) = h(\rho)g(\phi).
\end{equation}

\begin{figure}[t]
\centering\includegraphics[width=1\columnwidth]{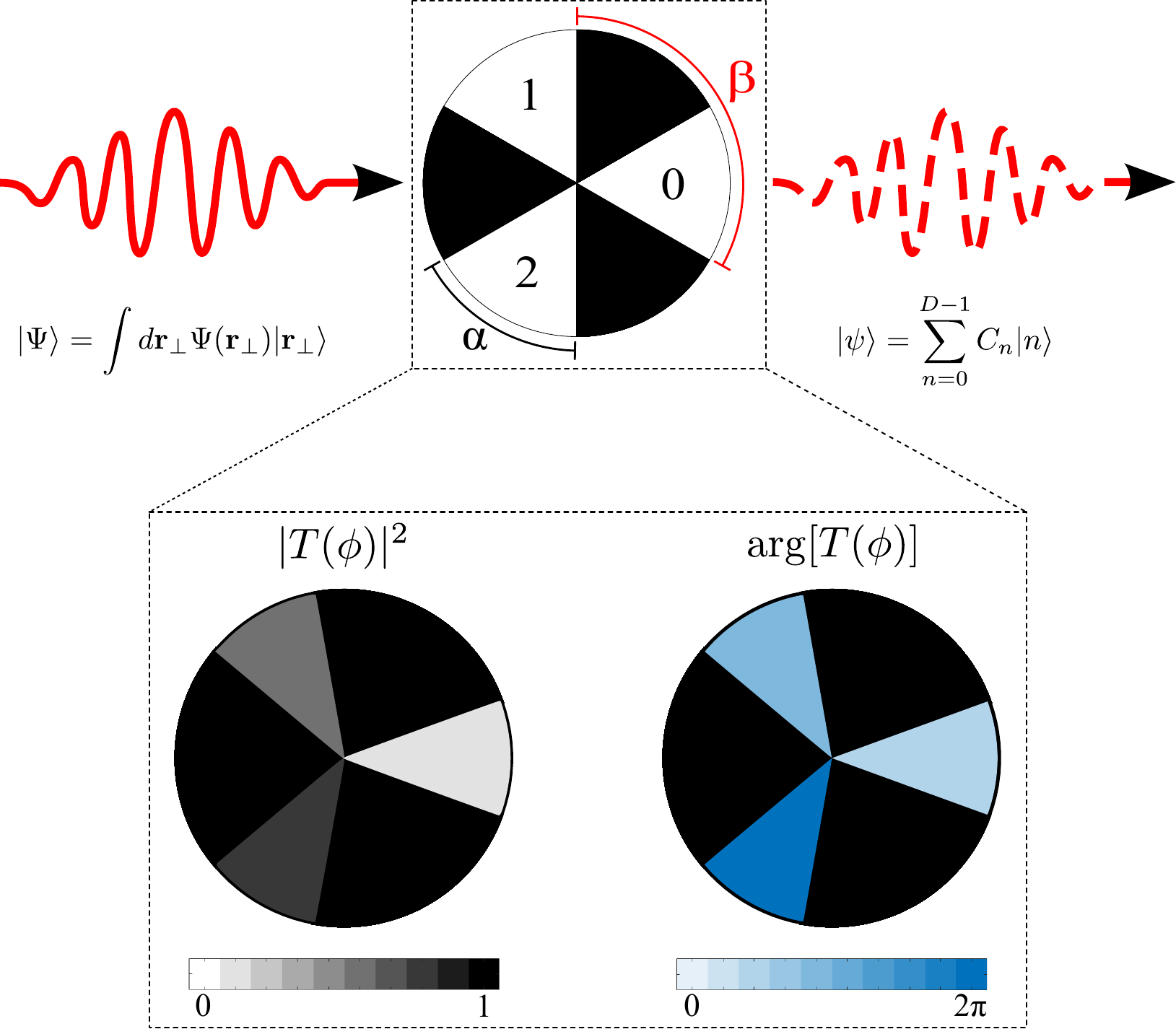}
    \caption{Schematic of the modulation of the spatial profile of a single photon by an array of $D=3$ angular slits of width $\alpha$ and period $\beta$. The angular position coordinate is discretized, generating an angular qudit state. The inset illustrates how the complex transmission function of the aperture $T(\phi)$ given by Eq.~(\ref{eq:aperture}) with $\alpha=2\pi/9$ can be implemented in practice by combining amplitude ($c_n\rightarrow|c_n|$) and phase  ($c_n\rightarrow e^{i\arg c_n}$) modulation masks. }
    \label{fig:FendaAngularEx}
\end{figure}

Suppose now that these photons are directed into an aperture whose complex transmission function is given by
\begin{equation} \label{eq:aperture}
    T(\phi) =\sum_{n=0}^{D-1} c_n \Pi\left(\frac{\phi - n \beta}{\alpha}\right),
\end{equation}
where $D\geq 2$, $\Pi \left( \frac{\phi - n \beta}{\alpha} \right)$ is a rectangle function of width $\alpha$ centered in $n\beta$, and $\{c_n\}$ are complex coefficients that satisfy $|c_n| \in [0,1]$. This aperture consists of an array of $D$ angular slits of width $\alpha$ and period $\beta$ (labeled by the index $n$), which are symmetrically distributed in the transverse plane (i.e., $\beta = 2\pi/D$) and modulated by a complex amplitude. Figure~\ref{fig:FendaAngularEx} illustrates the scenario described above considering $D=3$; the inset shows a way to manipulate the coefficients $\{c_n\}$, which will be discussed later. The spatial profile of the photon passing through this aperture will be proportional to $T(\phi) \Psi(\rho, \phi)$. Since the input amplitude [Eq.~(\ref{eq:separability})] is separable in $\rho$ and $\phi$, and the aperture depends only on the azimuthal coordinate, we can integrate the output amplitude in $\rho$, obtaining
\begin{equation}\label{eq:output}
    \psi(\phi) = N T(\phi) g(\phi),
\end{equation}
where $N$ is a normalization constant that absorbs the integration. Thus, the  state of the photon transmitted through the aperture [Eq.~(\ref{eq:aperture})] will be written as
\begin{align}
    |\psi\rangle &=  \int_{-\pi}^{\pi}  d\phi \, \psi(\phi) \, |\phi\rangle \nonumber \\
    &= \int_{-\pi}^{\pi}  d\phi \, \left[N \sum_{n=0}^{D-1} c_n  \Pi\left(\dfrac{\phi - n \beta}{\alpha}\right)  g(\phi) \right]|\phi\rangle.
\end{align}
Assuming that the function $g(\phi)$ does not vary considerably in the range $[-\alpha/2,\alpha/2]$, it can be treated as a constant [$\approx g(n\beta)$] within each slit and thus can be dropped from the integral, leaving us with
\begin{align} 
|\psi\rangle &= N \sum_{n=0}^{D-1} c_n \,g(n \beta)\int_{-\pi}^{\pi}   d\phi \, \Pi\left( \dfrac{\phi - n \beta}{\alpha} \right)   |\phi\rangle \nonumber\\
        &= \sum_{n=0}^{D-1} C_n |n\rangle,
\label{eq:AngularQudit}
\end{align}
where
\begin{equation}\label{eq:Cn}
    C_n \equiv N {c_n} \, g(n\beta),
\end{equation}
and
\begin{equation}\label{eq:SlitState}
    |n\rangle \equiv \int_{-\pi}^{\pi}  d\phi \, \Pi \left( \dfrac{\phi - n \beta}{\alpha} \right) |\phi\rangle.
\end{equation}

Equation~(\ref{eq:AngularQudit}) shows us that the state of the photon, initially defined in continuous variables [Eq.~(\ref{eq:estadoinicial})], after transmission through the aperture $T(\phi)$ [Eq.~(\ref{eq:aperture})], can be written as a discrete sum of $D$ states of angular position defined by the slits, as sketched in Fig.~\ref{fig:FendaAngularEx}. The ket $|n\rangle$ in Eq.~(\ref{eq:SlitState}) represents the state of a photon transmitted through the $n$th slit of the aperture and will be referred to as the angular slit state. It is straightforward to show that, since $\beta>\alpha$, these states are orthonormal, i.e.,
\begin{align}
\langle n'|n\rangle &= \int_{-\pi}^{\pi} d \phi \, \Pi (\phi) \, \Pi \left( \phi - \dfrac{(n' - n)\beta}{\alpha} \right) \nonumber\\
&= \Lambda \left( \dfrac{(n' - n)\beta}{\alpha} \right)=\delta_{n,n'},
\end{align}
where $\Lambda(x)=1-|x|$ if $|x|<1$ [$\Lambda(x)=0$ if $|x|\geq 1$] is the triangle function. Therefore, 
Eq.~(\ref{eq:AngularQudit}) represents the state of a qudit, specifically a photonic angular qudit, where the angular slit states $\{|n\rangle\}_{n=0}^{D-1}$ form an orthonormal and canonical basis that spans a $D$-dimensional Hilbert space.

\begin{figure*}[t]
    \centering
    \includegraphics[width=0.75\textwidth]{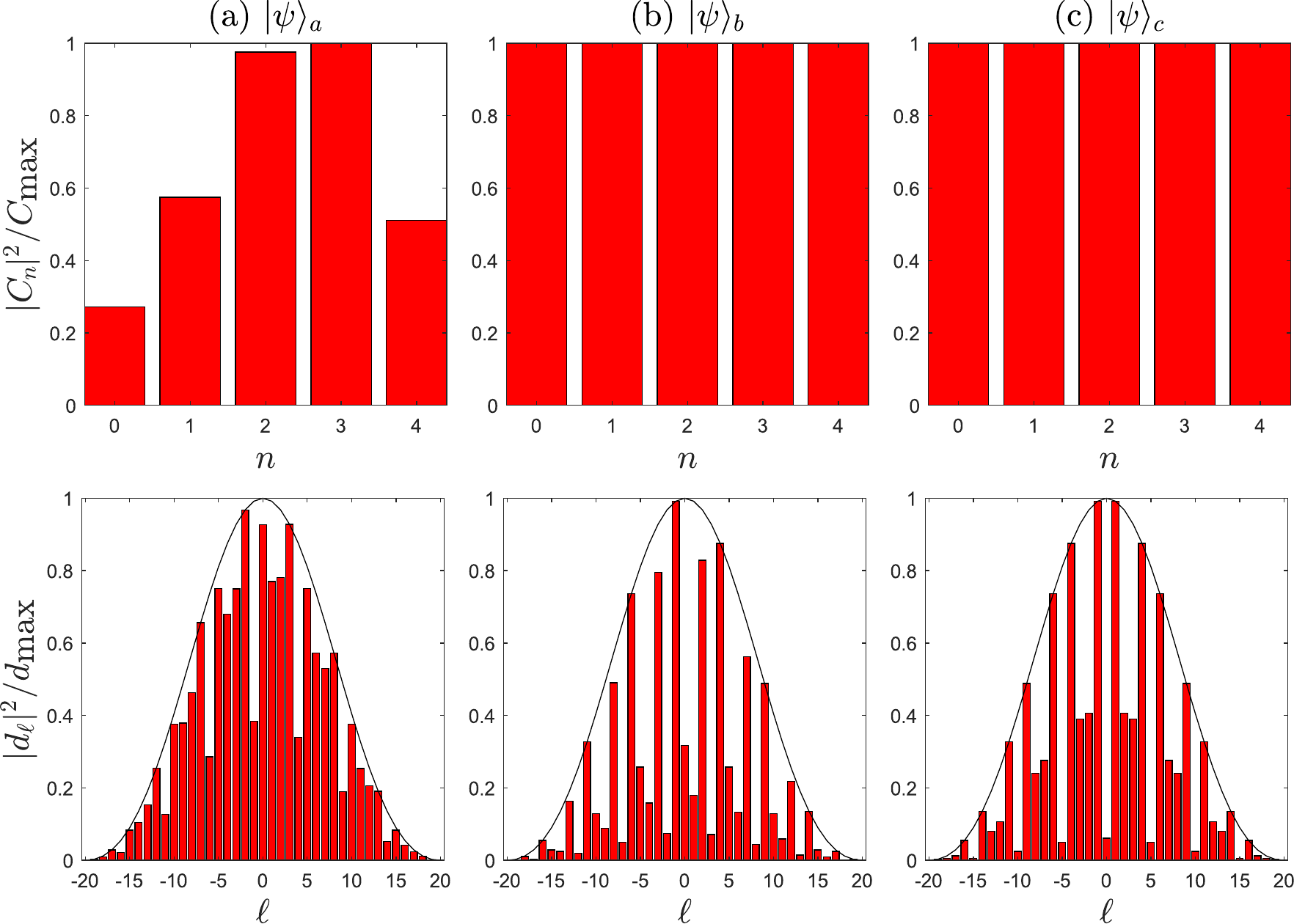}
    \caption{Five-dimensional angular qudit states $|\psi\rangle_{a,b,c}$ (see text). Normalized distributions of the squared magnitudes of the coefficients in the slit basis (top) and OAM basis (bottom). The black curve corresponds to the envelope $\textrm{sinc}^2(\ell \alpha/2)$. }
    \label{fig:D5Ex}
\end{figure*}

\begin{figure*}[t]
    \centering
    \includegraphics[width=0.75\textwidth]{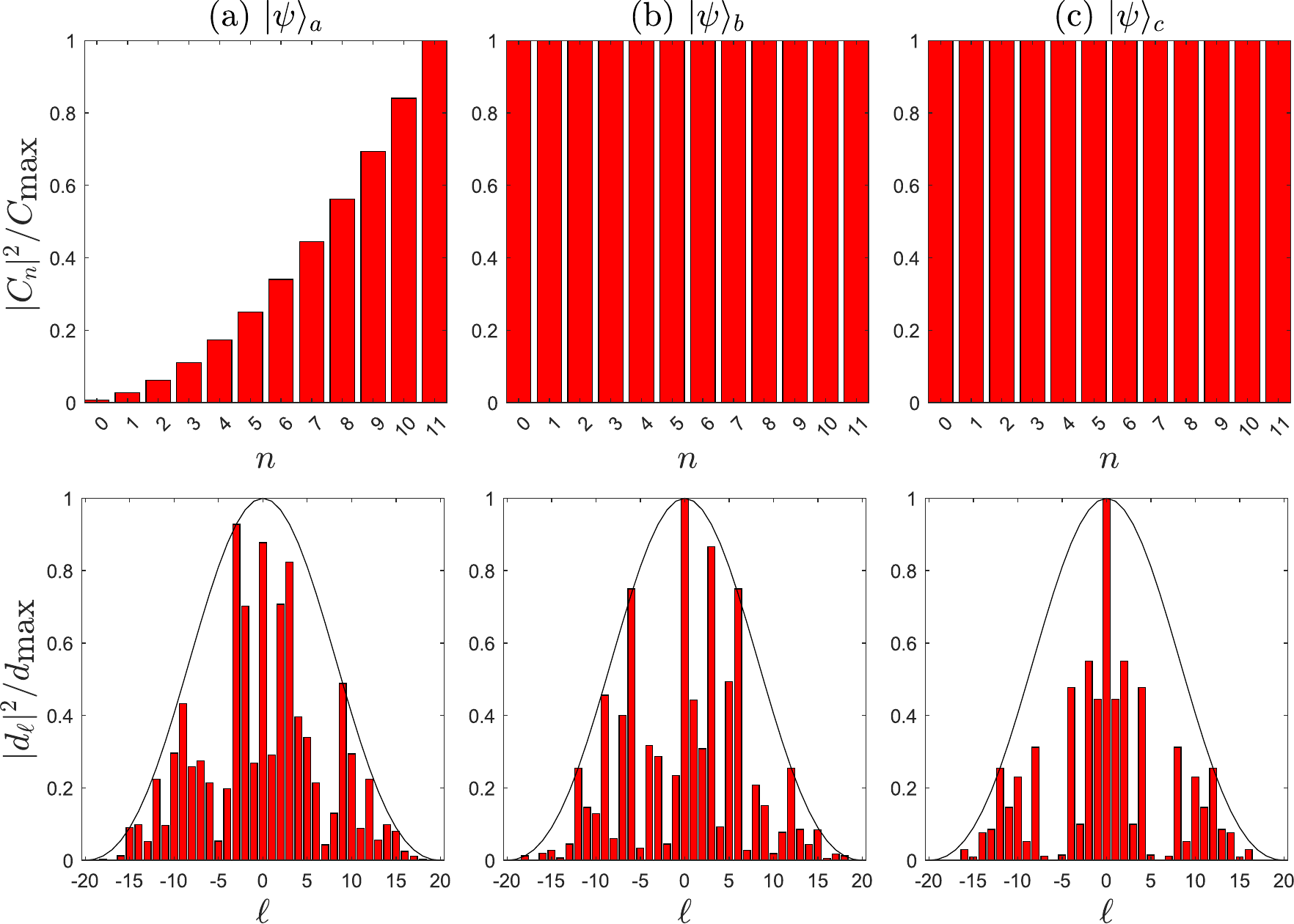}
    \caption{Twelve-dimensional angular qudit states $|\psi\rangle_{a,b,c}$ (see text). Normalized distributions of the squared magnitudes of the coefficients in the slit basis (top) and OAM basis (bottom). The black curve corresponds to the envelope $\textrm{sinc}^2(\ell \alpha/2)$.}
    \label{fig:D12Ex}
\end{figure*}

The angular qudit state is characterized by the complex coefficients $C_n$ given by Eq.~(\ref{eq:Cn}). These coefficients, in turn, depend on the spatial profile of the incident photons at the positions $\phi = n\beta$ of the aperture [$g(n\beta)$], and on the complex amplitudes $c_n$ that modulate each slit of the transmission function $T(\phi)$ [Eq.~(\ref{eq:aperture})]. Therefore, by manipulating both elements, we can control the preparation of this state. A simple approach is to produce a spatial profile close to a plane wave, so that the preparation will be controlled only by manipulating the transmission function: the coefficients $\{c_n\}$ can be generated by combining masks that modulate their magnitudes $\{|c_n|\}$ and phases $\{\arg(c_n)\}$, as sketched in the inset of Fig.~\ref{fig:FendaAngularEx} for $D=3$. In practice, this can be made dynamically using programmable spatial modulators (SLMs), for instance, by combining amplitude-only and phase-only SLMs \cite{Lima2011}. Alternatively, one can also use a single phase-only SLM with a mask given by the product of a binary array of $D$ angular slits with a specific diffraction grating applied to each slit. This method has been demonstrated for photonic qudits encoded in position and linear momentum \cite{Prosser2013}, and may be extended to angular qudits; a deeper analysis of this subject is beyond the scope of the present work.

\subsection{Angular qudit state in the OAM basis}

The angular position and orbital angular momentum of light are conjugate variables, i.e., there is a Fourier relationship between them. Due to the periodic nature of $\phi$, its conjugate variable $\ell$ is discrete \cite{Jack2008,Yao2006}. For single-photon states, this leads to the definition of the conjugate bases $\{|\phi\rangle\}$ and $\{|\ell\rangle\}$, whose vectors are related as follows:
\begin{align}
    |\phi\rangle &= \dfrac{1}{\sqrt{2\pi}} \sum_{\ell = -\infty}^{\infty} e^{-i\ell \phi} |\ell\rangle,\label{eq:ketphi} \\
    |\ell\rangle &= \dfrac{1}{\sqrt{2 \pi}} \int_{-\infty}^{\infty}d\phi \, e^{i \ell \phi} |\phi\rangle. \label{eq:ketl}
\end{align}

By substituting Eq.~(\ref{eq:ketphi}) into Eq.~(\ref{eq:SlitState}) and performing simple algebraic manipulations, we obtain the angular slit state in the OAM basis as
\begin{align}  \label{eq:ketnOAM}
    |n\rangle &= \dfrac{\alpha}{\sqrt{2 \pi}} \sum_{\ell = -\infty}^{\infty} e^{-i \ell n \beta} {\textup{sinc}}( \ell\alpha /2) |\ell\rangle, 
\end{align}
where $\textrm{sinc}(x)=\sin(x)/x$ is the Fourier transform of a rectangle function. Thus, the angular qudit state in Eq.~(\ref{eq:AngularQudit}) will be given in the OAM basis by
\begin{equation}\label{eq:AngularQuditOAM}
    |\psi\rangle = \sum_{\ell = -\infty}^{\infty} d_{\ell}|\ell\rangle,
\end{equation}
where
\begin{equation}\label{eq:dl}
    d_\ell = \dfrac{\alpha}{\sqrt{2 \pi}} {\textup{sinc}}( \ell\alpha /2) \sum_{n=0}^{D-1} C_n e^{-i \ell n \beta} .
\end{equation}
This result shows that the angular qudit in the OAM basis is defined within an infinite-dimensional discrete space. However, as we will see in the next section, projective measurements in this basis will be restricted to $D$-dimensional subspaces of this degree of freedom.

\subsubsection{Examples}\label{subsec:ExQudit}

To illustrate the results presented so far, let us exemplify the angular qudit states considering their representations in both the slit basis [Eq.~(\ref{eq:AngularQudit})] and the OAM basis [Eq.~(\ref{eq:AngularQuditOAM})]. For this, we generate $D$-dimensional states through a normalized vector of $D$ complex numbers that represent the coefficients $C_n$ of the state in the slit basis. From it, we can generate the corresponding state in the OAM basis by determining the coefficients $d_\ell$ using Eq.~(\ref{eq:dl}); we consider an aperture of $D$ angular slits with period $\beta = 2\pi/D$ and width $\alpha=\pi/10$. With these data we will calculate and plot the normalized distributions of $|C_n|^2/C_\textrm{max}$, where $C_\textrm{max} = \max_n \{|C_n|^2\}$, and $|d_\ell|^2/d_\textrm{max} $, where $d_\textrm{max} = \max_\ell \{|d_\ell|^2\}/ \textrm{sinc}^2\left(\ell_\textrm{max} \alpha /2 \right)$, and $\ell_\textrm{max}$ is the value of $\ell$ with the largest probability amplitude.

First, considering $D=5$, we generate a random state,  $|\psi\rangle_a$, and two uniform states (i.e., $|C_n| = 1/\sqrt{5}$ $\forall$ $n$): $|\psi\rangle_b$ with random relative phases and $|\psi\rangle_c$ with fixed relative phases given by $\arg\{C_n\}=\pi\times(0,0.8,-0.8,0.8,0)$. Figure~\ref{fig:D5Ex} shows the distributions in the slit basis (top) and OAM basis (bottom). Similarly, we also generate three states for $D=12$, with the results shown in Fig.~\ref{fig:D12Ex}:  $|\psi\rangle_a$ has random relative phases with the magnitudes $|C_n|=0.039223(n+1)$. The states $|\psi\rangle_b$ and $|\psi\rangle_c$ are uniform (i.e., $|C_n| = 1/\sqrt{12}$ $\forall$ $n$); the former has random relative phases while the latter has fixed relative phases given by $\arg\{C_n\}=\frac{\pi}{5}\times(0,2,1,2,4,-2,-2,4,2,1,2,0)$.

The superposition of angular position states, represented by the slits, generates interference in the distribution of OAM modes, analogous to what is observed in a multislit array that restricts the linear position of a light beam and causes interference between the components of the transverse linear momentum. The origin of this effect can be analyzed from the shape of the coefficients $d_\ell$ in Eq.~(\ref{eq:dl}): the squared modulus $|d_\ell|^2$ gives rise to interference due to the summation, and it is modulated by the envelope $\textrm{sinc}^2(\ell \alpha/2)$. The shape of this envelope is due to the angular diffraction of a photon at a slit, given by a rectangular function in cylindrical coordinates. The effect of this diffraction affects the measurement efficiency in the OAM basis, as we will see in the next section. In the bottom row of Figs.~\ref{fig:D5Ex} and \ref{fig:D12Ex}, the black curves represent the central maximum of $\textrm{sinc}^2 (\ell \alpha/2)$; the other maxima were omitted because they will not be useful in the ptychographic method.

\section{Ptychographic estimation of photonic \mbox{angular} qudit states}
\label{sec:PtyAngQudit}

The measurement steps of the ptychographic method for estimating unknown pure quantum states are summarized in the diagram below:
\begin{equation}
|\psi\rangle \longrightarrow \hat{P}_j|\psi\rangle \longrightarrow\left|\langle f_n|\hat{P}_j |\psi\rangle\right|^2.
\end{equation}
 In this section, we will discuss how to implement these steps for angular qudits and demonstrate the application of the method for the states shown in Figs.~\ref{fig:D5Ex} and \ref{fig:D12Ex}.

\subsection{Implementation of the ptychographic projectors} 
\label{subsec:PtyProjectors}

As described in Section~\ref{sec:Ptychography}, the first step of the ptychographic method is the projection of the unknown quantum state of a qudit, $|\psi\rangle$, into $\lceil D/2 \rceil$-dimensional subspaces with some overlap between them. For this purpose, a set of $J$ rank-$\lceil D/2 \rceil$ projectors, $\{\hat{P}_j\}_{j=0}^{J-1}$, has been introduced. These ptychographic projectors, defined in Eq.~(\ref{eq:PtyProjectors}), are diagonal in the canonical basis of the state space. They act as binary filters that select $\lceil D/2 \rceil$ components of $|\psi\rangle$ and suppress the others.

In the case of angular qudits, whose canonical basis is given by the angular slit states, projectors of this form will be implemented simply by binary spatial filters that block $D-\lceil D/2 \rceil$ slits. According to the shape of the angular mode, defined by a rectangular function in the azimuthal coordinate [see Eq.~(\ref{eq:SlitState})], the transmission function of the  filter that implement the $j$th projector will be given by
\begin{align}   \label{eq:SpatialFilters}
T_j(\phi) & = \sum_{k=0}^{\lceil D/2\rceil-1}\Pi\left[\frac{\phi-(k\oplus s_j)\beta}{\alpha'}\right].
\end{align}
This is an array of $\lceil D/2\rceil$ angular slits of width $\alpha'\geq\alpha$ that have maximal transmittance and are located at the same positions as the angular modes of the qudit state. In an experimental implementation, the width of these slits, $\alpha'$, can be slightly larger than the angular modes to ensure that, when these modes are imaged into the filters, they will be transmitted without risk of loss.

\begin{figure}[t]
\centering\includegraphics[width=.85\columnwidth]{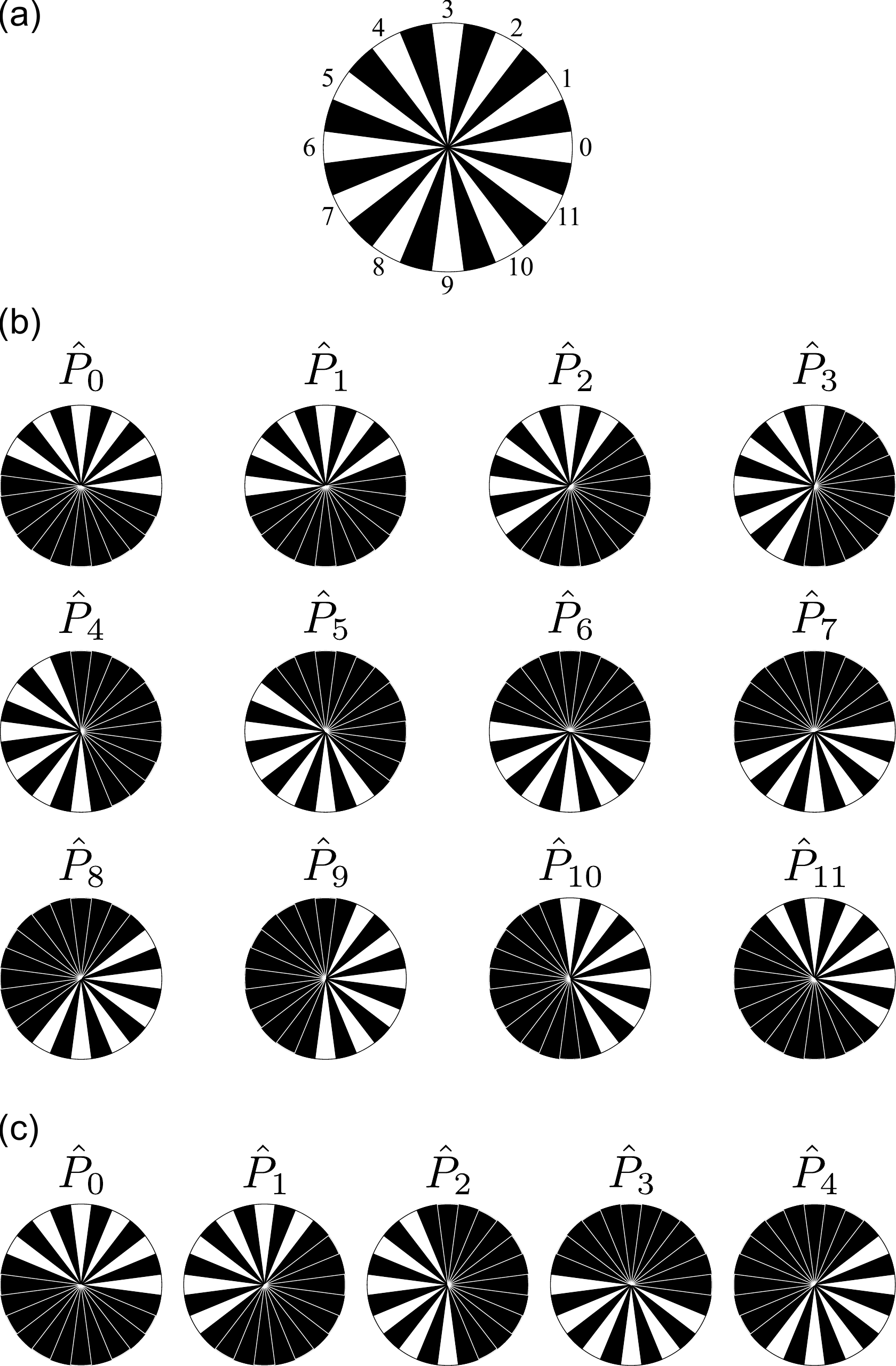}
    \caption{(a) Map locating the positions of the angular spatial modes in a twelve-dimensional angular qudit. Family of $J=12$ (b) and $J=5$ (c) binary spatial filters, generated from Eq.~(\ref{eq:SpatialFilters}) with $\alpha'=\pi/12$, that carry out the rank-$6$ ptychographic projections $\{\hat{P}_j\}$ given by Eq.~(\ref{eq:PtyProjectors}).}
    \label{fig:D12PtyProj}
\end{figure}

To illustrate this discussion, let us consider $D=12$.  Figure~\ref{fig:D12PtyProj} shows sets of spatial filters that carry out rank-$6$ projections. Figure~\ref{fig:D12PtyProj}(a) represents a map that locates the position of the angular spatial modes. The family of $J=12$ ptychographic projectors with $\mathbf{s}_D=(0,\ldots,D-1)$ is shown in Fig.~\ref{fig:D12PtyProj}(b). Figure~\ref{fig:D12PtyProj}(c) shows the filters corresponding to the family of $J=5$ projectors with $\mathbf{s}_5=(0,2,4,6,8)$.


\subsection{Measurement in the Fourier basis}

After each ptychographic projection, the method proceeds with the measurement in the Fourier basis given by Eq.~(\ref{eq:FourierBasis}). For the angular qudits, the states of this basis, $\{|f_n\rangle\}$, are uniform superpositions of the angular slit states. Using Eq.~(\ref{eq:ketnOAM}), we can rewrite them in the OAM basis as
\begin{align}
|f_n\rangle &= \sum_{\ell=-\infty}^{\infty}\Phi_{\ell n}|\ell\rangle,
\end{align}
where
\begin{align}   \label{eq:Phi_ln}
\Phi_{\ell n} &= K\textrm{sinc}\left(\frac{\ell\alpha}{2}\right)
\frac{\sin\left(\frac{D\beta(n-\ell)}{2}\right)}{\sin\left(\frac{\beta(n-\ell)}{2}\right)}
e^{i\beta(n-\ell)(D-1)/2},
\end{align}
with $K=\alpha/\sqrt{2\pi D}$. The derivation of $\Phi_{\ell n}$ is shown in Appendix~\ref{app:Phi_ln}. These coefficients show more explicitly the interference effect that occurs in the distribution of OAM modes, as discussed earlier.
 
\begin{figure*}[t]
    \centering
    \includegraphics[width=.9\textwidth]{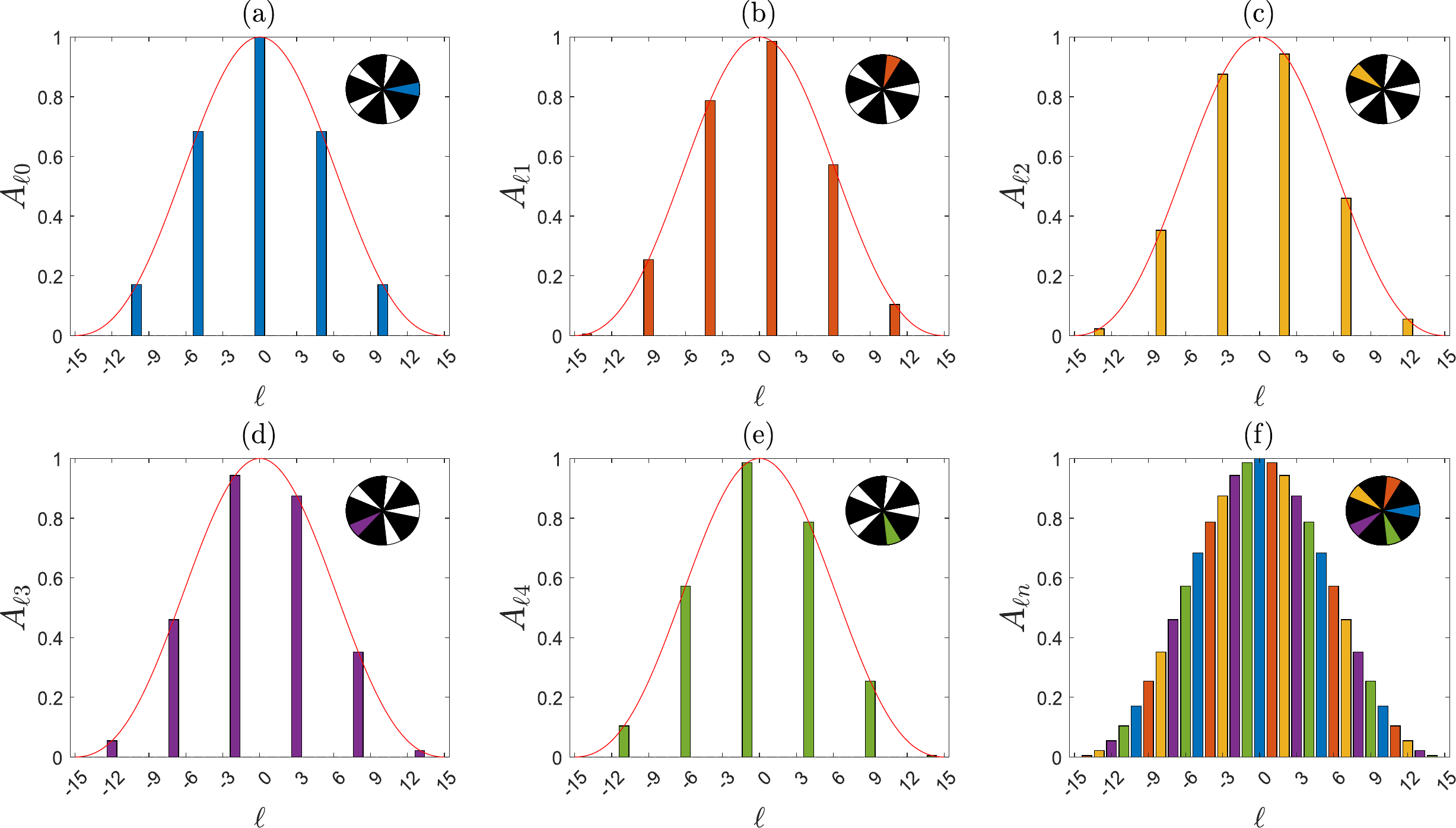}
    \caption{Normalized probability distribution [Eq.~(\ref{eq:AellnDef})] for the OAM modes ($\ell$) associated with the Fourier-basis state (a) $|f_0\rangle$, (b) $|f_1\rangle$, (c) $|f_2\rangle$, (d) $|f_3\rangle$, and (e) $|f_4\rangle$ of a five-dimensional angular qudit. (f) Overlay of the previous plots. The insets highlight the angular slit state over which QFT was applied, and the red lines represent the sinc-shaped diffraction envelope.} 
    \label{fig:Aell5D}
\end{figure*}

The most interesting aspect of writing the Fourier basis in the OAM representation is that the QFT of each angular slit state generates specific values of $\ell$, distinct from the other states, as a result of the orthogonality $\langle f_{n'}|f_n\rangle=\langle n'|\hat{\mathcal{F}}_D^\dag\hat{\mathcal{F}}_D|n\rangle=\langle n'|n\rangle=\delta_{n,n'}$. To illustrate this, let us consider 
\begin{align}  \label{eq:AellnDef}
A_{\ell n} &= \left|\frac{\Phi_{\ell n}}{\Phi_{00}}\right|^2,
\end{align}
which represents the probability distribution for the $\ell$ modes of the state $|f_n\rangle$, normalized by the maximum achievable probability $|\langle 0|f_0\rangle|^2$. Taking as an example $D=5$ angular slits with $\alpha = \pi/20$, Figs.~\ref{fig:Aell5D}(a)-\ref{fig:Aell5D}(e) show this distribution for $\{|f_n\rangle\}_{n=0}^{4}$; the insets of each plot highlight the angular slit state over which the QFT was applied, and the red lines represent the $\textrm{sinc}^2(\ell\alpha/2)$ envelope. Figure~\ref{fig:Aell5D}(f) shows the plots of $\{A_{\ell n}\}_{n=0}^{4}$ overlaid. For each state $|f_n\rangle$ measured in the OAM basis, it can be observed that 
\begin{equation} \label{eq:Aelln}
A_{\ell n}\neq 0\iff \ell=n+xD,
\end{equation}
where $x$ is an integer number. This result shows that there is a one-to-one correspondence between an OAM state $|\ell\rangle$ and a Fourier-basis state $|f_n\rangle$. Thus, by selecting $D$ OAM modes, each corresponding to a state $|f_n\rangle$, the projective measurement in the Fourier basis can be carried out in a $D$-dimensional subspace of the infinite-dimensional OAM space.

From Eq.~(\ref{eq:Aelln}), there are infinite possibilities for the selection of these $D$ OAM modes. For instance, a trivial choice consists of taking $\ell=0,\ldots,D-1$ which correspond to the states $|f_0\rangle,\ldots,|f_{D-1}\rangle$, respectively. However, for the sake of efficiency, which will be discussed below, it is more appropriate to choose modes encompassing $\ell=0$ and those around it. Specifically, for odd $D$, we have 
\begin{align}   \label{eq:Dodd}
\ell(n) &= \left(n\oplus\frac{D-1}{2}\right)-\frac{D-1}{2},
\end{align}
while for even $D$,
\begin{align}   \label{eq:Deven}
\ell(n) &= \left(n\oplus\frac{D}{2}\right)-\frac{D}{2},
\end{align}
where $\oplus$ denotes addition modulo $D$. For example, if $D=5$, the states $|f_0\rangle,\ldots,|f_4\rangle$ correspond to the modes $\ell=0,1,2,-2,-1$ (see Fig.~\ref{fig:Aell5D}); if $D=4$, $|f_0\rangle,\ldots,|f_3\rangle$ correspond to $\ell=0,1,-2,-1$.

The measurement in the Fourier basis performed in this way is affected by a loss of efficiency, as the OAM modes with $\ell\neq 0$ will have reduced amplitudes due to angular diffraction, as can be seen in Fig.~\ref{fig:Aell5D}. Thus, the choice of modes guided by Eqs.~(\ref{eq:Dodd}) and (\ref{eq:Deven}) is the optimal one to minimize this effect. An additional step is to apply compensation to the experimental data. In the ptychographic method, each post-projection state $|\psi_j\rangle$ is measured in the Fourier basis, generating a photon count rate $C_{j\ell}\propto|\langle\ell|\psi_j\rangle|^2$. This rate can be compensated during postprocessing by
\begin{equation} \label{eq:compesation}
\tilde{C}_{j\ell}=\frac{C_{j\ell}}{\textrm{sinc}^2(\ell\alpha/2)}. 
\end{equation}

\subsection{Numerical simulations}

\begin{figure*}[htbp]
    \centering
    \includegraphics[width=.79\textwidth]{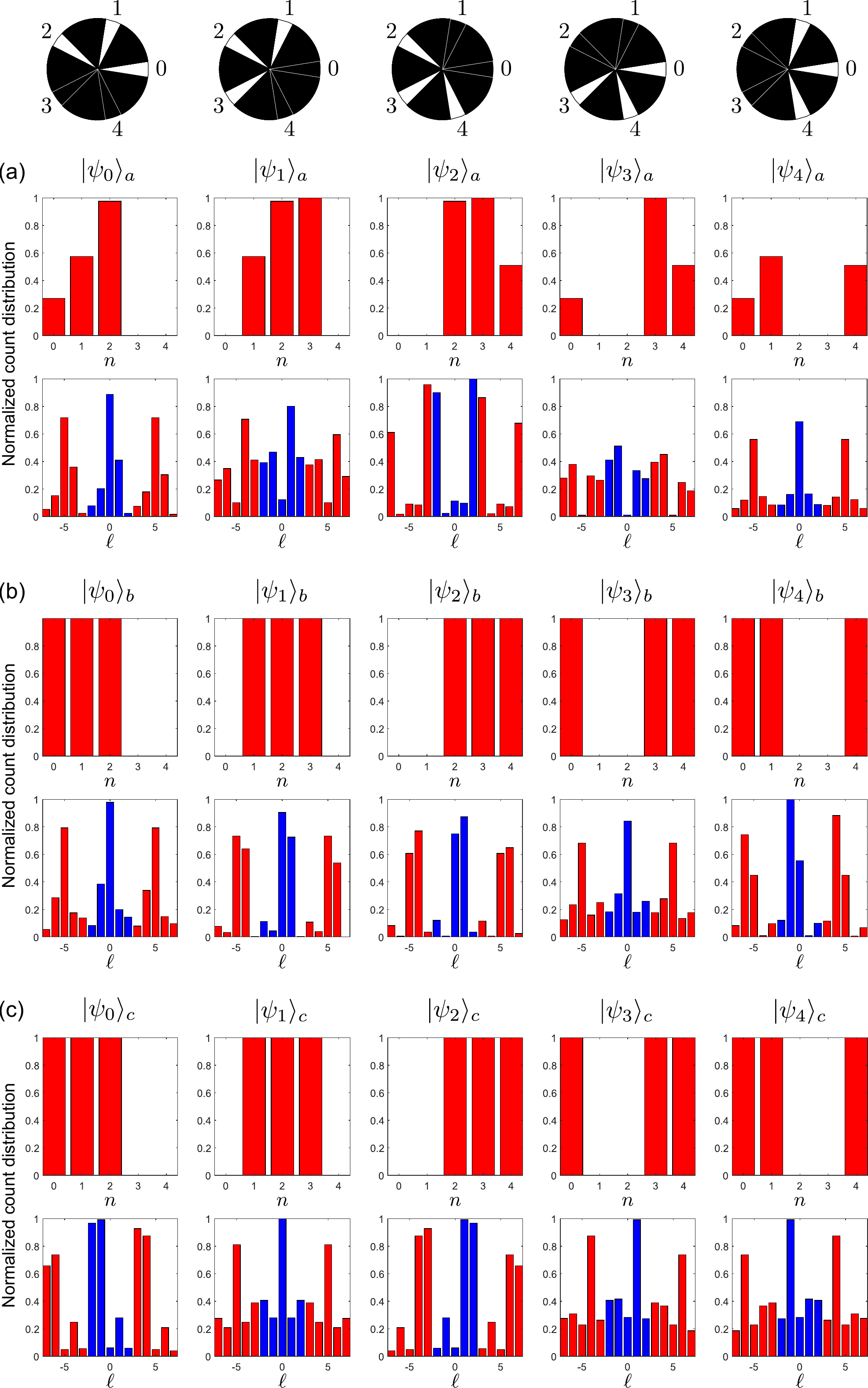}
    \caption{Normalized count distributions in the bases $\{|n\rangle\}$ and $\{|\ell\rangle\}$ for the states $|\psi\rangle_{a,b,c}$ shown in Fig.~\ref{fig:D5Ex} after ptychographic projections: (a) $\hat{P}_j|\psi\rangle_a$, (b) $\hat{P}_j|\psi\rangle_b$, and (c) $\hat{P}_j|\psi\rangle_c$ ($j=0,\ldots,4$). Insets at the top show the spatial filters that implement $\{\hat{P}_j\}$. The blue bars correspond to the ptychographic data.} 
    \label{fig:PtySym5D}
\end{figure*}

\begin{figure*}[t]
    \centering
    \includegraphics[width=.82\textwidth]{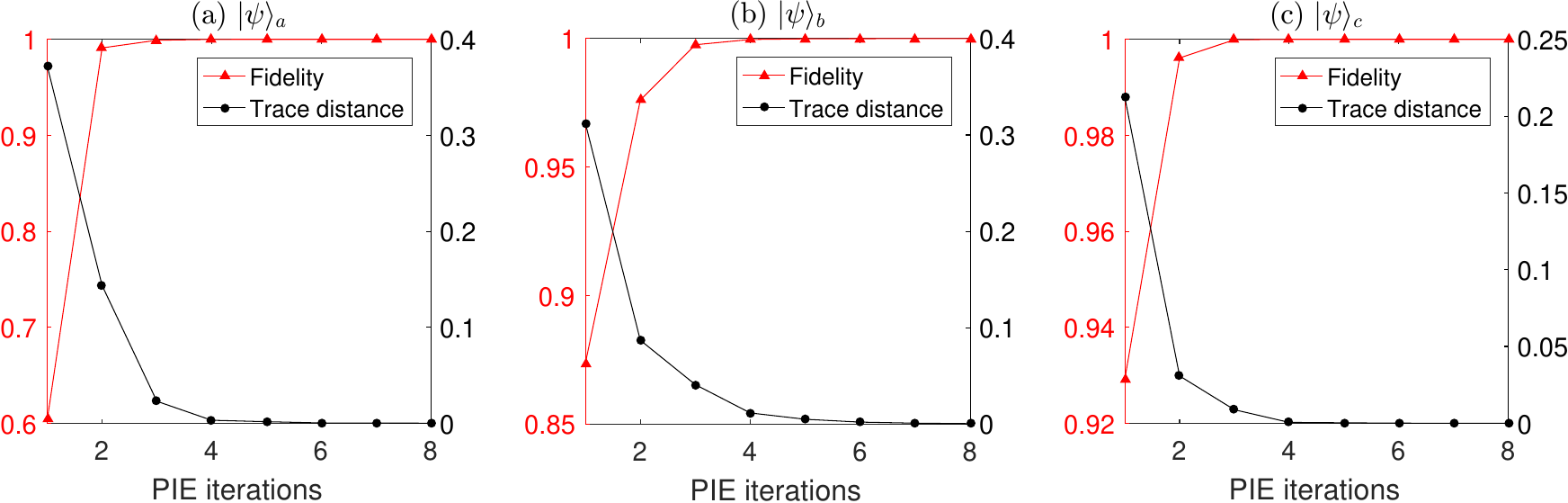}
    \caption{Progression of the PIE algorithm for estimating the five-dimensional angular qudit states $|\psi\rangle_{a,b,c}$ shown in Fig.~\ref{fig:D5Ex}. } 
    \label{fig:Pty5}
\end{figure*}

\begin{figure*}[h!]
    \centering
    \includegraphics[width=.82\textwidth]{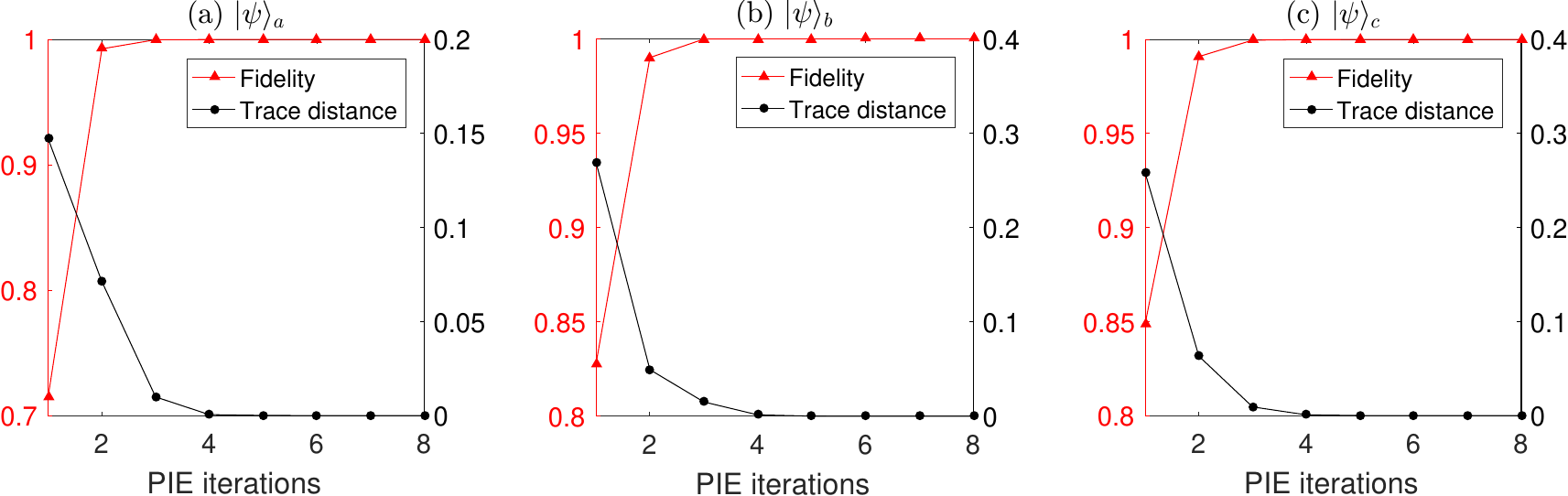}
	 \vskip 5mm
    \includegraphics[width=.82\textwidth]{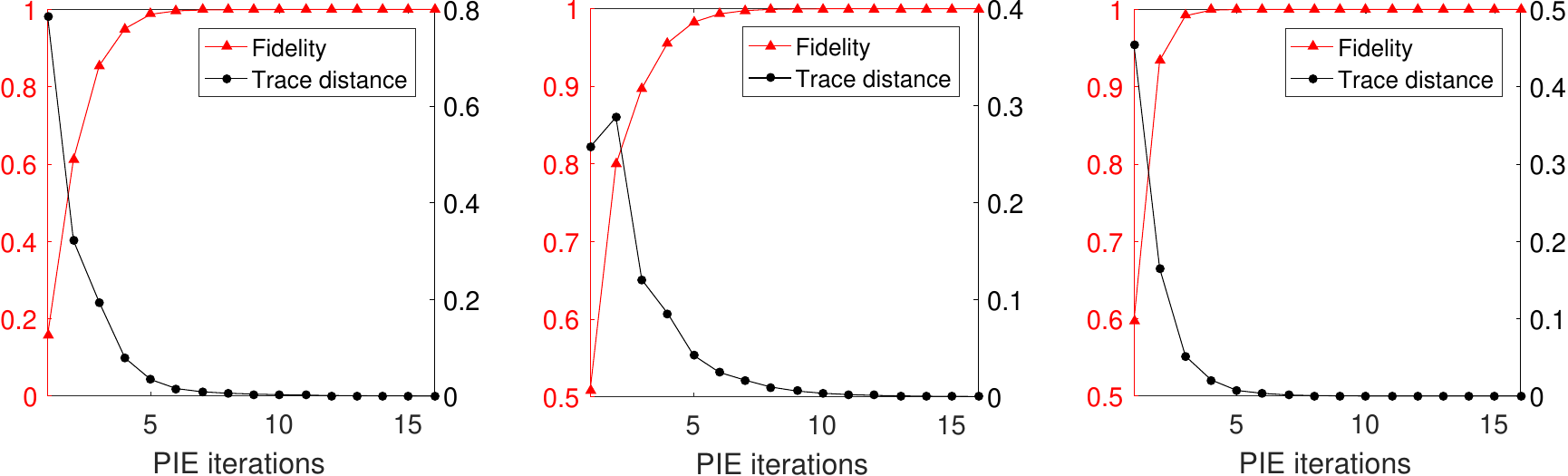}
    \caption{Progression of the PIE algorithm for estimating the twelve-dimensional angular qudit states $|\psi\rangle_{a,b,c}$ shown in Fig.~\ref{fig:D12Ex}. Estimations performed with $J=12$ (top) and $J=5$ (bottom) ptychographic projectors. } 
    \label{fig:Pty12}
\end{figure*}

To demonstrate the effectiveness of the ptychographic method described above, we apply it to estimate the five- and twelve-dimensional angular qudit states $|\psi\rangle_{a,b,c}$ shown in Figs.~\ref{fig:D5Ex} and \ref{fig:D12Ex}, respectively. First, for $D=5$, we consider a set of $5$ rank-$3$ projectors of the form given by Eq.~(\ref{eq:PtyProjectors}), with $\mathbf{s}_5=(0,\ldots,4)$. The binary spatial filters that implement these projectors are illustrated at the top of Fig.~\ref{fig:PtySym5D}. For the states generated after the ptychographic projections, namely $|\psi_j\rangle_{a,b,c}=\hat{P}_j|\psi\rangle_{a,b,c}$, Figs.~\ref{fig:PtySym5D}(a)-\ref{fig:PtySym5D}(c) show their normalized count distributions in the bases $\{|n\rangle\}$ (first row) and $\{|\ell\rangle\}$ (second row). It can be seen that the application of each binary filter completely blocks the transmission of some slits, altering the output states. This is reflected in the interference that defines the distribution of the OAM modes; the blue bars in these plots correspond to the ptychographic data, which are obtained from the modes given by Eq.~(\ref{eq:Dodd}). These specific count rates were compensated according to Eq.~(\ref{eq:compesation}) and, together with the set $\{\hat{P}_j\}$, were fed into the PIE algorithm described in Section~\ref{sec:Ptychography}. To run the algorithm, we used the feedback parameter $\eta=1.5$ and a fixed number of $8$ PIE iterations. In addition to calculating the trace distance [Eq.~(\ref{eq:trace_distance})] to verify the convergence of the algorithm, we also compute 
\begin{equation}\label{eq:fidelity}
F = |\langle\Upsilon_{\textrm{updt}}|\psi\rangle|^2,
\end{equation}
which is the fidelity between the target state ($|\psi\rangle_{a,b,c}$) and the normalized updated estimate ($|\Upsilon_{\textrm{updt}}\rangle$) provided at each iteration. The results obtained are shown in Fig.~\ref{fig:Pty5}, where we observe that the estimations were successful. Overall, fewer than $8$ iterations were required for the algorithm to converge, with the fidelity reaching its maximum value ($F=1$).

For the states $|\psi\rangle_{a,b,c}$ with $D=12$ (Fig.~\ref{fig:D12Ex}), we performed the same process described above using either $J=12$ or $J=5$ rank-$6$ ptychographic projectors; their corresponding spatial filters are those in Fig.~\ref{fig:D12PtyProj}(a) and \ref{fig:D12PtyProj}(b), respectively. The OAM modes associated with the measurement in the Fourier basis of the post-projection states were selected according to Eq.~(\ref{eq:Deven}). Figure~\ref{fig:Pty12} shows the results obtained with $J=12$ (top) and $J=5$ (bottom). In the first case, we again used $8$ PIE iterations, while in the second case, we used $16$. With fewer projectors, we will have a smaller set of ptychographic data, resulting in the need for more iterations to achieve the desired convergence and fidelity. In both cases, the estimations were successful, achieving $F=1$.

\subsection{Assessing experimental feasibility}

\begin{figure}[t]
    \centering
    \includegraphics[width=1\columnwidth]{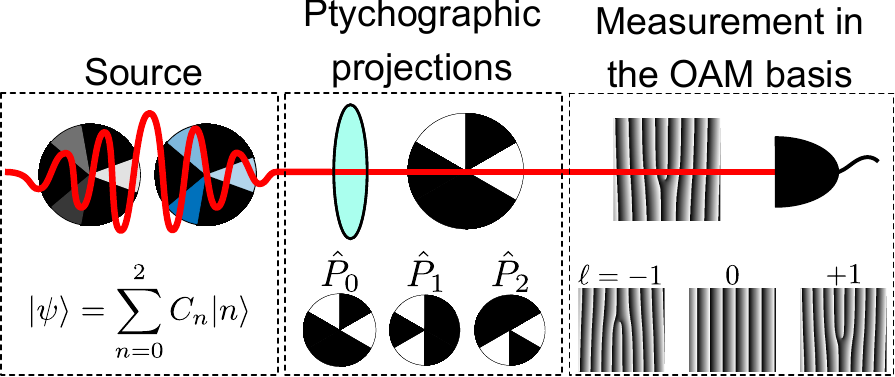}
    \caption{Schematic representation of the stages of the ptychographic method applied for a photonic angular qutrit. The angular modes are imaged from the source to the binary spatial filters, and the state emerging from each projection is measured in the OAM basis (see text for more details).  } 
    \label{fig:setup}
\end{figure}

An important aspect to address is the experimental feasibility of the ptychographic method for photonic angular qudits. Figure~\ref{fig:setup} shows a general scheme for this purpose. A qudit pure state is prepared by a source according to the description provided in Section~\ref{subsec:AngularQudit2}. Next, the angular modes of the qudit are imaged from the source to the plane where the ptychographic projections will be carried out by the binary spatial filters. Finally, the state emerging from each projection is projectively measured in the OAM basis. 

The spatial filters given by Eq.~(\ref{eq:SpatialFilters}) (see also Fig.~\ref{fig:D12PtyProj}) could be implemented dynamically through an SLM that modulates only the amplitude of the incident single-photon field. Generating the filters as described in Section~\ref{subsec:PtyProjectors} will mitigate the losses produced by misalignments in the imaging system, although some unavoidable loss by diffraction is expected from the pixel structure of the device. The measurement in the OAM basis for the post-projection states could be performed in several ways \cite{Yao2011}.  Here, we mention two of them. First, the simplest approach involves using forked diffraction holograms \cite{Bazhenov1992,Heckenberg1992}. These holograms convert the target OAM mode into a Gaussian mode, which can then be coupled to a single-mode fiber and detected. Despite its simplicity, this technique allows the measurement of only one mode at a time. Additionally, its efficiency decreases as the number of modes involved in the experiment increases. A more efficient approach involves applying unitary transformations that convert a set of OAM modes into a set of tilted plane waves. By creating copies of the converted modes, they can be almost perfectly discriminated by their lateral positions at a focal plane of a lens \cite{Berkhout2010,OSullivan2012,Mirhosseini2013}. This method allows the simultaneous measurement of OAM modes with different values of $\ell$. Reference~\cite{Mirhosseini2013} reports the measurement of modes from $\ell=-12$ to $\ell=12$, which would enable ptychographic estimation of $25$-dimensional angular qudit states. 

Using less efficient approaches in this two-step process will require larger ensembles of identically prepared photonic angular qudits to achieve fidelities comparable to those achieved with more efficient approaches.

\section{Conclusion}
\label{sec:Conclusion}

We theoretically described the implementation of the ptychographic method for estimating pure quantum states of $D$-dimensional qudits encoded in the angular position and OAM of single photons. In the first step of the measurement process, we have shown that the partially overlapping projections are performed using binary spatial filters that select $\lceil D/2\rceil$ angular spatial modes and block the others. In the second step, we demonstrated that the postselection of $D$ OAM modes compatible with the QFT of the angular modes corresponds to the projective measurement in the Fourier basis. Additionally, we provided the optimal way to select the $D$ OAM modes among the infinite ones. These measurement steps were put to the test through simulations. The results obtained demonstrated the effectiveness of the method by perfectly estimating angular qudit states in dimensions $D=5$ and $D=12$.

Photonic angular qudits are generated in a simple way by discretizing the spatial profile of a photon in cylindrical coordinates. Several optical tools are available for manipulating and detecting the degrees of freedom that characterize this encoding (angular position and OAM). In this sense, they can be a valuable resource for applications in quantum information processing \cite{Wang2022} and quantum communications \cite{Mirhosseini2015} in high-dimensional state spaces. Although many challenges remain in developing such applications \cite{Dunlop2017}, they will benefit from the availability of state estimation methods in the future. Given that the current technology enables the preparation of angular qudit states with high purity, the ptychographic method may become particularly useful, as it is less costly than standard tomographic schemes in terms of experimental and computational resources.

\appendix
\section{Amplitudes of the Fourier-basis states in the OAM representation}  \label{app:Phi_ln}

The angular qudit states in the OAM basis are given by Eq.~(\ref{eq:ketnOAM}). By substituting it into Eq.~(\ref{eq:FourierBasis}), we obtain the OAM representation of the Fourier-basis states as
\begin{equation}   \label{eq:fn_app}
|f_n\rangle=K \sum^{\infty}_{\ell=-\infty} \textrm{sinc} \left(\frac{\ell\alpha}{2}\right)\left[ \sum^{D-1}_{k=0}e^{ik\beta(n-\ell)}\right] |\ell\rangle,
\end{equation}
where $K=\alpha/\sqrt{2\pi D}$. Using the definition of a finite geometric series, the sum in square brackets will be written as
\begin{align}
[\cdots]&= \frac{1-e^{iD\beta(n-\ell)}}{1-e^{i\beta(n-\ell)}} \nonumber\\
&= \frac{e^{-iD\beta(n-\ell)/2} - e^{iD\beta(n-\ell)/2} }{e^{-i\beta(n-\ell)/2}-e^{i\beta(n-\ell)/2}}e^{i\beta(n-\ell)(D-1)/2}  \nonumber\\
&= 
\frac{\sin\left(\frac{D\beta(n-\ell)}{2}\right)}{\sin\left(\frac{\beta(n-\ell)}{2}\right)}
e^{i\beta(n-\ell)(D-1)/2}.
\end{align}
By substituting this expression into Eq.~(\ref{eq:fn_app}), we arrive at the definition of the amplitudes $\Phi_{\ell n}$ given by Eq.~(\ref{eq:Phi_ln}).

\vskip 5mm

\noindent\textbf{Acknowledgments.} This work was supported by CNPq INCT-IQ (465469/2014-0), CNPq (422300/2021-7), and CNPq (303212/2022-5). A. M. C. acknowledges support from CAPES.




\end{document}